\documentclass[12pt]{article}
\usepackage[margin=1in]{geometry}
\usepackage{amsmath,amssymb,amsthm}
\usepackage{natbib}
\usepackage{hyperref}
\usepackage{setspace}

\newtheorem{theorem}{Theorem}
\newtheorem{proposition}{Proposition}
\theoremstyle{definition}
\newtheorem*{definition*}{Definition}
\newtheorem*{example*}{Example}
\newtheorem*{remark*}{Remark}

\newcommand{\ba}{\begin{align}}
\newcommand{\ea}{\end{align}}
\newcommand{\be}{\begin{equation}}
\newcommand{\ee}{\end{equation}}

\newcommand{\Core}{\operatorname{Core}}
\newcommand{\Aut}{\operatorname{Aut}}
\newcommand{\ED}{\operatorname{ED}}

\newcommand{\cE}{\mathcal{E}} 
 
\newcommand{\cG}{\mathcal{G}} 
 
\newcommand{\cI}{\mathcal{I}}

\newcommand{\cV}{\mathcal{V}}

\newcommand{\E}{{\mathbb E}}

\newcommand{\R}{{\mathbb R}}

\newcommand{\nn}{\nonumber}

\title{Stable Allocations, Unstable Payment Paths:\\The Shapley Value and Voluntary Coalition Formation}
\author{Tongseok Lim\\Purdue University\\\texttt{lim336@purdue.edu}}
\begin{document}
\maketitle

\begin{abstract}
The random-order interpretation of the Shapley value specifies both a terminal allocation and a payment path: each entrant is paid her marginal contribution at the time of entry. We embed this path in a finite dynamic entry problem in which a selected player may accept immediately or defer and allow another outsider to enter first. A flow is fully voluntary when immediate acceptance is optimal at every decision state. Although this criterion is defined through optimal continuation values, backward induction yields an equivalent system of linear inequalities in full-consent continuation values. The distinction between allocation and path is sharp in convex games: the Shapley allocation lies in the core, but the Shapley flow is fully voluntary only in additive games. Nevertheless, every efficient allocation admits a budget-balanced fully voluntary implementation when signed payments are available. We select a canonical implementation by minimizing distance from the Shapley flow, derive a reduced quadratic program for symmetric cardinality games, and study equal residual sharing, which selects the terminal allocation endogenously. With two player types, the selected allocation is an affine combination of the Shapley and equal-division allocations. Examples illustrate the effects of payment restrictions and a possible conflict between full voluntariness and coalitional stability.
\end{abstract}


\section{Introduction}
\label{sec:introduction}

The Shapley value is usually viewed as an allocation rule. Its familiar
random-order interpretation, however, also describes a payment process.
Players enter one at a time, and each entrant receives the increase in
coalition value caused by her arrival. Averaging these payments over all
entry orders gives the Shapley allocation \citep{Shapley1953}.

This interpretation implicitly assumes that players are willing to
take their assigned positions in the entry order. However, if a player's marginal contribution is larger after more partners
have entered, she may prefer to reject an early offer and wait for a
later position. The random-order formula still defines an allocation,
but the payment rule embedded in that formula does not support the
coalition-formation path used to calculate it.

The issue is especially clear in convex cooperative games. Convexity
means that marginal contributions increase as the coalition grows. It
also gives the Shapley allocation a strong cooperative stability
property: every marginal vector belongs to the core, and therefore the
Shapley value belongs to the core \citep{Shapley1971,Ichiishi1981}. Yet
the same increase in marginal contributions can make late entry more
attractive under contemporaneous marginal-contribution payments.
Convexity can therefore stabilize the final allocation while
destabilizing the payment path that generates it.

We separate these two objects. A \emph{payment flow} specifies what each
player receives when the coalition changes from $S$ to
$S\cup\{j\}$. Edge-by-edge budget balance requires total payments on
that transition to equal the value it creates. The usual random-order
interpretation is a special case, which we call the \emph{Shapley
flow}: the entrant receives the entire marginal contribution and all
other players receive zero. A general flow may instead divide the
transition surplus between the entrant and incumbents. Expected payments
along a uniformly random path determine the flow's terminal allocation.

We evaluate a flow in a finite dynamic entry problem. At each coalition, nature selects an outsider uniformly. A selected player may accept and enter immediately or defer the current opportunity, in which case another outsider is selected uniformly and enters first. The original player remains outside and may face the same choice at a later state. We define her optimal continuation value by backward induction, taking immediate acceptance by the other players as given. A flow is fully voluntary if immediate acceptance is optimal for every player at every decision state. We then show that this sequential criterion is equivalent to a system of linear inequalities in continuation values generated by universal immediate acceptance. This equivalence preserves the tractability of the original flow problem.

The first result characterizes when the Shapley flow is fully voluntary. A
player compares her marginal contribution now with the expected
marginal contribution at her eventual entry coalition. Under
submodularity, deferral cannot improve this payment. Under
supermodularity, a strict increase in marginal contribution creates a
profitable deferral. Combining this observation with the classical core
result for convex games yields the paper's main contrast. In every
nonadditive convex game, the Shapley allocation is core-stable, but the
Shapley flow is not fully voluntary. Among convex games, contemporaneous
marginal-contribution payments satisfy both requirements only in
additive games.

The failure concerns the Shapley payment path, not the Shapley
allocation. Fix any efficient target allocation $x$. For every
coalition, assign its worth among its members and pay changes in these
coalition-specific accounts as the coalition expands. The payments
telescope, making each player indifferent across entry dates, and the
terminal accounts can be chosen to equal $x$. Thus every efficient
allocation, including the Shapley allocation of a convex game, admits a
budget-balanced fully voluntary implementation when signed payments
are available.

This existence result leaves many possible implementations for a given allocation target. We select
the voluntary flow closest to the Shapley flow in Euclidean distance.
The criterion treats marginal-contribution pricing as the benchmark and
measures the payment adjustment required to remove holdout while
preserving the target allocation. The selected flow is unique,
respects player relabeling, and is invariant to a positive change in
the unit of account. For symmetric cardinality games, the original
flow problem reduces to a quadratic program with $n-1$ incumbent
transfers. The active participation constraints identify the coalition
sizes at which waiting incentives require an adjustment.

We then study a more restrictive institution that also selects the
terminal allocation. Under \emph{equal residual sharing}, any part of
a transition surplus not paid to the entrant is divided equally among
incumbents. We choose the voluntary equal-residual flow closest to the
Shapley flow and call its terminal allocation the equal-residual value.
The rule agrees with Shapley whenever the Shapley flow is already
fully voluntary. Otherwise, the payment restriction and the consent
constraints jointly determine how far the allocation moves away from
marginal-contribution pricing.

When the game has two player types, symmetry and efficiency imply that
type-symmetric efficient allocations lie on an affine line. The
equal-residual value can then be written as a game-dependent affine
combination of the Shapley and equal-division allocations. In the
three-player glove game, the signed-payment rule places weight $3/4$ on
Shapley and $1/4$ on equal division. Requiring nonnegative individual
payments reduces the Shapley weight to $13/23$. These weights are not
chosen in advance. They are outcomes of the payment-design problem.

Equal residual sharing need not preserve the core. In a convex
unanimity game where players 1 and 2 generate the entire surplus and
player 3 is null, the equal-residual value assigns a positive payoff to
player 3 and leaves the productive pair with less than its coalition
worth. The example shows that compensation supporting a payment process
can conflict with coalitional outside options. For two-type convex
games, this conflict can be summarized by a lower bound on the Shapley
weight required for core membership.

\subsection{Related literature}

The paper builds first on the Shapley value and the theory of convex
games. \citet{Shapley1953} introduced the value and its random-order
formula, and \citet{Weber1988} placed random-order values in a broader
class of probabilistic values. \citet{Shapley1971} and
\citet{Ichiishi1981} established the connection between convexity,
permutation marginal vectors, and the core. We retain the uniform
random-order benchmark but separate two objects that coincide in its
usual interpretation: the terminal allocation and the transition
payments made along the realized order.

A related literature provides noncooperative foundations and
implementation mechanisms for the Shapley value. \citet{Gul1989}
derives the value as a limit of efficient equilibria in a dynamic
bargaining model, \citet{HartMasColell1996} obtain it from an
$n$-person bargaining procedure, and \citet{PerezCastrilloWettstein2001}
implement it through a bidding-and-proposal mechanism. Those papers
design strategic procedures whose equilibrium payoffs coincide with
the Shapley value. Our question is different. We take the transition
payments embedded in the random-order formula as the benchmark and ask
whether immediate acceptance is sequentially optimal along that payment path. When it is not, we redesign the path while preserving the target allocation whenever desired.

The flow formulation also connects to graph-based representations of
cooperative games. \citet{SternTettenhorst2019} use combinatorial Hodge
decomposition on the coalition graph to decompose a game into
player-specific component games and recover the Shapley value at the
grand coalition; their construction also admits a least-squares
interpretation. Our use of the coalition lattice is complementary.
Rather than decomposing the characteristic function, we optimize a
vector of transition payments subject to edge-by-edge budget balance,
entry incentives, and, when imposed, a terminal-allocation constraint.
Thus the closest-flow criterion is a constrained payment-design problem,
not a Hodge decomposition of the game.

The paper is also related to allocation schemes defined for every
coalition. \citet{Sprumont1990} introduced population-monotonic
allocation schemes, under which an incumbent's assigned payoff does not
fall when a coalition expands. Our exact flows are changes in
coalition-specific accounts, so nonnegative exact flows are closely
related to population monotonicity. The incentive studied here runs in
the opposite direction along an edge: we ask whether an outsider
prefers to enter now rather than preserve the possibility of entering
later. Moreover, general voluntary flows in our model need not be exact
or nonnegative.

A further connection is to cooperative solutions that combine
participation and equality. \citet{DuttaRay1989} select egalitarian
allocations subject to participation constraints, while subsequent work
studies constrained egalitarianism, population solidarity, and
combinations of the Shapley and equal-division rules
\citep{DuttaRay1991,VanDenBrinkEtAl2016,VanDenBrinkFunaki2025}. In our
construction, equal treatment is imposed edge by edge on the residual paid
to incumbents rather than directly on the terminal allocation. The
resulting allocation is generated by the payment process, and the
weight placed on marginalism versus equality is selected endogenously
by the constrained projection.

Finally, the paper relates to dynamic coalition formation. Sequential
and dynamic models study how protocols and continuation values shape
coalitional outcomes \citep{Bloch1996,KonishiRay2003,RayVohra2015}.
Our protocol is intentionally narrower: it does not determine an endogenous coalition structure or model general bargaining, renegotiation, permanent withdrawal, or farsighted deviations. Players are committed to eventual participation in the grand coalition, but a selected outsider may defer the current entry opportunity. The model isolates whether immediate acceptance is sequentially optimal along an announced random-order payment path.

\subsection{Organization}

Section~\ref{sec:flows} introduces payment flows and the allocations
they induce. Section~\ref{sec:model} formulates the dynamic entry problem, establishes its linear characterization, and analyzes the Shapley flow. Section~\ref{sec:implementation}
contrasts core stability with path stability and develops voluntary
implementation of a prescribed target, including the symmetric
cardinality case. Section~\ref{sec:equalresidual} studies equal residual
sharing and the endogenous allocation it selects. The final section
concludes. Proofs and calculations are collected in the appendix.

\section{Payment flows and induced allocations}
\label{sec:flows}

Let $N=\{1,\ldots,n\}$ be a finite set of players. A transferable-utility
coalition game is a function
\[
 v:2^N\to\R,
 \qquad
 v(\varnothing)=0.
\]
Let $\cG_N$ denote the class of such games. The number $v(S)$ is the
worth coalition $S$ can generate, and $v(N)$ is the surplus to be
allocated if the grand coalition forms. For $i\notin S$, player $i$'s
marginal contribution to $S$ is
\begin{equation}
 \Delta_i v(S):=v(S\cup\{i\})-v(S).
 \label{eq:marginal-contribution}
\end{equation}

Let $\sigma$ be a permutation of $N$, interpreted as an entry order,
and let
\[
 S_i^\sigma:=\{j\in N:\sigma(j)<\sigma(i)\}
\]
be the set of players preceding $i$. The Shapley value is
\begin{equation}
 \phi_i(v) = \frac1{n!}\sum_\sigma \Delta_i v(S_i^\sigma).
 \label{eqn:shapley}
\end{equation}
Thus player $i$ receives her average marginal contribution over all
entry orders. The allocation is efficient:
\[
 \sum_{i\in N}\phi_i(v)=v(N).
\]

Formula \eqref{eqn:shapley} can be read as a payment rule. For a given
entry order, player $i$ is paid $\Delta_i v(S_i^\sigma)$ when she joins.
The formula averages these path payments. Our analysis allows the same
transition surplus to be distributed differently while retaining the
random-order structure.

A \emph{forward edge} is a pair $(S,S\cup\{j\})$ with $j\notin S$.
Let
\begin{equation}
 \cE := \{(S,S\cup\{j\}):S\subseteq N,\ j\notin S\}
 \label{eq:forward-edges}
\end{equation}
be the set of all forward edges. A payment flow for player $i$ is a
function $f_i:\cE\to\R$. A flow profile
$\mathbf f=(f_i)_{i\in N}$ specifies the payment to every player at
every possible coalition expansion. The payment $f_i(S,S\cup\{j\})$
is received by player $i$ when entrant $j$ joins coalition $S$.
Therefore, the payment recipient need not be the entrant.

We impose edge-by-edge budget balance:
\begin{equation}
 \sum_{i\in N}f_i(S,S\cup\{j\})
 =
 \Delta_jv(S)
 \qquad(S\subseteq N,\ j\notin S).
 \label{eq:budget}
\end{equation}
The payments made when $j$ enters must exhaust the value created by
that transition. Unless stated otherwise, individual payments may be
negative. This flexibility is useful for separating the existence of
an implementation from additional restrictions such as limited
liability.

The random-order interpretation of the classical Shapley value
corresponds to a particular flow.

\begin{definition*}[Shapley flow]
The Shapley flow $\partial v=(\partial_i v)_{i\in N}$ pays the full
transition surplus to the entrant:
\begin{equation}
 \partial_i v(S,S\cup\{j\})
 =
 \begin{cases}
 \Delta_jv(S),&i=j,\\
 0,&i\neq j.
 \end{cases}
 \label{eq:Shapley-flow}
\end{equation}
\end{definition*}

A general flow induces an allocation by averaging each player's total
payment over all entry orders.

\begin{definition*}[Flow-induced allocation]
For a flow profile $\mathbf f$, define
\begin{equation}
 \phi_{f_i}
 :=
 \frac1{n!}\sum_\sigma\sum_{j\in N}
 f_i(S_j^\sigma,S_j^\sigma\cup\{j\}),
 \qquad
 \phi_{\mathbf f}:=(\phi_{f_i})_{i\in N}.
 \label{eq:fvalue}
\end{equation}
We call $\phi_{\mathbf f}$ the allocation induced by $\mathbf f$.
\end{definition*}

The notation emphasizes the distinction at the center of the paper.
The flow $\mathbf f$ is a payment process on the coalition lattice,
whereas $\phi_{\mathbf f}$ is its expected terminal allocation. Many
flows can induce the same allocation.

Two immediate observations connect this framework to the classical
Shapley value. First,
\begin{equation}
 \phi_{\partial v}=\phi(v),
 \label{eq:Shapley-flow-recovery}
\end{equation}
because under $\partial v$ player $i$ is paid only when she enters, and
that payment equals $\Delta_i v(S_i^\sigma)$. Second, if $\mathbf f$
satisfies \eqref{eq:budget}, then $\phi_{\mathbf f}$ is efficient. For
each permutation path,
\begin{align*}
 \sum_{i\in N}\sum_{j\in N}
 f_i(S_j^\sigma,S_j^\sigma\cup\{j\})
 &=
 \sum_{j\in N}
 \bigl[v(S_j^\sigma\cup\{j\})-v(S_j^\sigma)\bigr]\\
 &=v(N)-v(\varnothing)=v(N).
\end{align*}
Averaging over permutations gives
\begin{equation}
 \sum_{i\in N}\phi_{f_i}=v(N).
 \label{eq:flow-efficiency}
\end{equation}

These observations also clarify what changing a payment flow can and
cannot do. Budget balance fixes total surplus on every transition and
ensures efficiency at the end, but it does not determine who is paid on
a given edge. That additional flexibility is what allows payment paths
to be redesigned in response to entry incentives.

\section{Payment paths and voluntary entry}
\label{sec:model}
For $S\subsetneq N$, let $m(S):=n-|S|$ denote the number of outsiders. Before coalition formation begins, the designer announces the complete state-contingent payment flow $\mathbf f$. Starting from a coalition $S$, nature selects one outsider uniformly. If selected player $i$ accepts, the transition $S\to S\cup\{i\}$ occurs and all payments $f_k(S,S\cup\{i\})$, $k\in N$, are made. If $i$ defers and at least one other outsider remains, nature selects one player uniformly from $N\setminus(S\cup\{i\})$; that player enters, the corresponding edge payments are made, and player $i$ remains outside. Player $i$ may again choose between acceptance and deferral if selected at a later state. When $i$ is the sole remaining outsider, deferral is no longer available and the final transition occurs.

The protocol treats rejection as a deferral of entry priority rather than withdrawal from the grand coalition. Equivalently, players are committed ex ante to eventual participation, but not to accepting every intermediate position offered by nature. To test whether universal immediate acceptance is sequentially rational, fix player $i$, hold fixed immediate acceptance by every other player, and allow $i$ to optimize dynamically over her own acceptance decisions.

\paragraph{Optimal continuation values.}
For a fixed flow $\mathbf f$, let $V_i(S)=V_i(S;\mathbf f)$ be player $i$'s optimal expected payment from state $S$ onward under this decision problem. Set $V_i(N)=0$. If $i\in S\subsetneq N$, player $i$ has already entered and has no further entry decision, so
\begin{equation}
V_i(S)=\frac{1}{m(S)}\sum_{j\notin S}
\left[f_i(S,S\cup\{j\})+V_i(S\cup\{j\})\right].
\label{eq:V-incumbent}
\end{equation}
If $i\notin S$ and $m(S)\ge2$, define the value of accepting immediately by
\begin{equation}
A_i(S;V):=f_i(S,S\cup\{i\})+V_i(S\cup\{i\}),
\label{eq:accept-value-V}
\end{equation}
and the value of deferring the current opportunity by
\begin{equation}
B_i(S;V):=\frac{1}{m(S)-1}
\sum_{j\in N\setminus(S\cup\{i\})}
\left[f_i(S,S\cup\{j\})+V_i(S\cup\{j\})\right].
\label{eq:defer-value-V}
\end{equation}
The first term in each summand is retained because a general flow may pay player $i$ when another outsider enters. The Bellman recursion is
\begin{equation}
V_i(S)=\frac{1}{m(S)}\max\{A_i(S;V),B_i(S;V)\}
+\frac{m(S)-1}{m(S)}B_i(S;V).
\label{eq:V-Bellman}
\end{equation}
With probability $1/m(S)$, player $i$ is selected and chooses between acceptance and deferral. Otherwise, another outsider enters, giving player $i$ the continuation value summarized by $B_i(S;V)$.

If $S=N\setminus\{i\}$, then $i$ is the sole outsider and
\begin{equation}
V_i(N\setminus\{i\})=f_i(N\setminus\{i\},N).
\label{eq:V-terminal}
\end{equation}
Equations \eqref{eq:V-incumbent}--\eqref{eq:V-terminal} define $V_i$ uniquely by backward induction on coalition size.

\begin{definition*}[Full voluntariness]
A payment flow $\mathbf f$ is \emph{fully voluntary} if immediate acceptance is optimal at every decision state:
\begin{equation}
A_i(S;V)\ge B_i(S;V)
\qquad
(i\notin S,\ m(S)\ge2).
\label{eq:dynamic-consent}
\end{equation}
Thus the all-accept strategy profile is a sequential best response for every player. We use ``voluntary'' and ``fully voluntary'' interchangeably below.
\end{definition*}

\paragraph{Full-consent continuation values.}
The dynamic definition appears nonlinear because the recursion contains a maximum. We now derive an equivalent linear representation using continuation values under universal immediate acceptance.

For $S\subsetneq N$, let $\cV_S
:= \{S\cup\{j\}:j\notin S\}$ and $\cV_N:=\varnothing$. 
A \emph{forward random walk} (FRW) is a Markov process
$X=(X_t)_{t\geq 0}$ on $2^N$ such that, conditional on
$X_t=S\subsetneq N$, the next state is selected uniformly from
$\cV_S$. The grand coalition is absorbing. Hence, starting from
$S$, the process reaches $N$ after exactly $n-|S|$ transitions. Since a realization of the FRW starting from
$X_0=\varnothing$ is equivalent to drawing a uniformly random permutation
of $N$, we have
\[
\phi_{f_i} = \E\left[ \sum_{t=1}^n f_i(X_{t-1},X_t) \right].
\]
Define player $i$'s continuation payoff by
\begin{equation}
U_i(S)=U_i(S;\mathbf f):=
\mathbb E^S\left[\sum_{t=1}^{n-|S|}f_i(X_{t-1},X_t)\right]
\label{eq:continuation}
\end{equation}
where $\E^S$ denotes expectation conditional on $X_0=S$. \eqref{eq:continuation} shows that $U_i(S)$ is linear in $f_i$. Moreover,
\begin{equation}
U_i(N)=0,
\qquad
U_i(S)=\frac1{m(S)}\sum_{j\notin S}
\left[f_i(S,S\cup\{j\})+U_i(S\cup\{j\})\right],
\label{eq:bellman}
\end{equation}
and at the empty coalition, $U_i(\varnothing;\mathbf f)=\phi_{f_i}$. 
For $i\notin S$ with $m(S)\ge2$, define
\begin{align}
A_i(S;U)&:=f_i(S,S\cup\{i\})+U_i(S\cup\{i\}),
\label{eq:accept-value-U}\\
B_i(S;U)&:=\frac{1}{m(S)-1}
\sum_{j\in N\setminus(S\cup\{i\})}
\left[f_i(S,S\cup\{j\})+U_i(S\cup\{j\})\right].
\label{eq:defer-value-U}
\end{align}
The recursion \eqref{eq:bellman} gives
\begin{equation}
U_i(S)=\frac1{m(S)}A_i(S;U)+\frac{m(S)-1}{m(S)}B_i(S;U).
\label{eq:U-convex-combination}
\end{equation}

\begin{proposition}[Dynamic foundation and linear characterization]
\label{prop:dynamic-equivalence}
For any payment flow $\mathbf f$, the following statements are equivalent:
\begin{enumerate}
\item $\mathbf f$ is fully voluntary in the sense of \eqref{eq:dynamic-consent};
\item $V_i(S)=U_i(S)$ for every player $i$ and coalition $S$;
\item for every $i\notin S$ with $m(S)\ge2$,
\begin{equation}
A_i(S;U)\ge B_i(S;U);
\label{eq:linear-dynamic-consent}
\end{equation}
\item for every $i\notin S$ with $m(S)\ge2$,
\begin{equation}
f_i(S,S\cup\{i\})+U_i(S\cup\{i\})\ge U_i(S).
\label{eq:consent}
\end{equation}
\end{enumerate}
Consequently, full voluntariness defines a closed convex polyhedron in the space of flows. 
\end{proposition}
Although the behavioral definition uses the optimal values $V_i$, thanks to Proposition \ref{prop:dynamic-equivalence}, the analysis below may use the equivalent linear constraints \eqref{eq:consent} for fully voluntary flows.

For $i\notin S$, define the linear consent gap
\begin{equation}
g_i(S;\mathbf f):=
f_i(S,S\cup\{i\})+U_i(S\cup\{i\})-U_i(S).
\label{eq:gap}
\end{equation}
By Proposition~\ref{prop:dynamic-equivalence}, $\mathbf f$ is fully voluntary if and only if $g_i(S;\mathbf f)\ge0$ at every decision state. At $S=N\setminus\{i\}$, the final transition is mandatory and the corresponding equality holds automatically.

\paragraph{Local flows and superharmonicity.}
We say that a flow $\mathbf f$ is \emph{local} if only the entrant and existing members can be paid on an edge:
\begin{equation}
f_i(S,S\cup\{j\})=0
\qquad\text{whenever }i\notin S\cup\{j\}.
\label{eq:locality}
\end{equation}
Locality simplifies the deferral value because an outsider receives no payment when another outsider enters.

\begin{proposition}[Consent as superharmonicity]
\label{prop:superharmonicity}
Suppose $\mathbf f$ is local. Then for $i\notin S$,
\begin{equation}
g_i(S;\mathbf f)
=(m(S)-1)U_i(S)-
\sum_{j\in N\setminus(S\cup\{i\})}U_i(S\cup\{j\}).
\label{eq:gap-identity}
\end{equation}
Consequently, $\mathbf f$ is fully voluntary if and only if
\begin{equation}
U_i(S)\ge
\frac{1}{m(S)-1}
\sum_{j\in N\setminus(S\cup\{i\})}U_i(S\cup\{j\})
\label{eq:superharmonicity}
\end{equation}
for every $i\notin S$ with $m(S)\ge2$.
\end{proposition}

Thus, under local payments, full voluntariness is equivalent to superharmonicity of each outsider's full-consent continuation value on the sublattice of coalitions excluding that player. In economic terms, the value of accepting now cannot be improved by passing the current opportunity and allowing a randomly selected outsider to enter first.

We next apply the dynamic criterion to marginal-contribution pricing, i.e., the Shapley flow $\partial v$. We say that a game $v$ is \emph{submodular} if
\begin{equation}
\Delta_i v(S)\ge\Delta_i v(T)
\qquad
\text{whenever }S\subseteq T\subseteq N\setminus\{i\},
\label{eq:submodular}
\end{equation}
and \emph{supermodular} if the reverse inequality holds. A game that is both submodular and supermodular is additive: $v(S) = \sum_{i \in S} w(i)$ for some function $w$ on $N$.

\begin{proposition}[When is marginal-contribution pricing fully voluntary?]
\label{prop:voluntary}
The Shapley flow $\partial v$ is fully voluntary if and only if
\begin{equation}
\Delta_i v(S)\ge \mathbb E^S[\Delta_i v(R_i^S)]
\qquad(i\notin S,\ m(S)\ge2),
\label{eq:exacttest}
\end{equation}
where $R_i^S$ is the random coalition immediately before $i$ enters when she defers at $S$ and all subsequent offers are accepted. Consequently, the Shapley flow is fully voluntary in every submodular game. In a supermodular game, any strict increase in marginal contribution along a nested pair of coalitions, that is, $\Delta_i v(S)<\Delta_i v(T)$ for some $i\in N$ and $S\subseteq T\subseteq N\setminus\{i\}$, creates a state at which immediate acceptance is not optimal; hence the Shapley flow is not fully voluntary.
\end{proposition}

The proposition has a sequential interpretation. Under the Shapley flow, a player who accepts receives her current marginal contribution and no subsequent payment. If she defers, she preserves the option to enter after additional players have joined. Submodularity makes this option unattractive, whereas complementarity can make it strictly valuable.

\section{Stable allocations and implementable payment paths}
\label{sec:implementation}

For an allocation $x=(x_1,\ldots,x_n)\in\R^n$ and a coalition
$S\subseteq N$, we write $x(S):=\sum_{i\in S}x_i$. The core of
$v\in\cG_N$ is defined by
\[
\Core(v)
:=
\left\{
x\in\R^n:
x(N)=v(N)
\text{ and }
x(S)\geq v(S)
\text{ for every }S\subseteq N
\right\}.
\]
The equality $x(N)=v(N)$ requires the entire worth of the grand
coalition to be allocated. The inequalities $x(S)\geq v(S)$ impose
coalitional rationality: the members of coalition $S$ receive, in
total, at least what they could obtain by acting on their own. Hence no
coalition can block a core allocation by leaving the grand coalition
and distributing $v(S)$ among its members.

Core membership is therefore a stability requirement on the terminal
allocation. It does not explain how the grand coalition forms or
whether players are willing to follow a proposed payment path. The dynamic condition in \eqref{eq:dynamic-consent}, equivalently its linear form \eqref{eq:consent}, addresses this distinct question.

A game is \emph{convex} if it is supermodular. Equivalently, each
player's marginal contribution is nondecreasing as the coalition of
other players expands:
\[
\Delta_i v(S)\leq\Delta_i v(T)
\qquad
\text{whenever }
S\subseteq T\subseteq N\setminus\{i\}.
\]
\citet{Shapley1971} showed that every permutation marginal vector of a
convex game belongs to its core. Since the Shapley value is the average
of these marginal vectors and the core is convex, it follows that
\[
\phi(v)\in\Core(v).
\]
Thus convexity gives the Shapley allocation a strong coalitional
stability property. At the same time, increasing marginal
contributions may make a player prefer a later position in the entry
order.

\begin{theorem}[Stable allocation, unstable payment path]
\label{thm:main}
Let $v$ be convex. Then the Shapley allocation belongs to the core. If
$v$ is nonadditive, however, the Shapley flow is not voluntary.
Consequently, among convex games, contemporaneous
marginal-contribution payments generate a Shapley allocation that is
both core-stable and supported by voluntary entry if and only if the
game is additive.
\end{theorem}

The first conclusion in Theorem~\ref{thm:main} is the classical core
result of \citet{Shapley1971}. The second conclusion concerns the
payment process and follows from Proposition~\ref{prop:voluntary}. In
a nonadditive convex game, at least one player's marginal contribution
rises as the coalition grows. If that player is offered entry before the relevant partners have joined, deferral has a strictly larger continuation value than immediate acceptance.

The theorem therefore does not say that the Shapley allocation is
unattainable. It says that the particular payment path implicit in the
random-order formula, namely the Shapley flow $\partial v$, does not
support voluntary entry. This distinction matters because transition
payments can be rearranged while preserving the same expected terminal
allocation.

\paragraph{Implementing a prescribed allocation.}
Fix an efficient target $x$, so $x(N)=v(N)$. For every coalition $S$,
choose accounts $y_i(S)$ such that
\begin{equation}\label{def:accounts}
y_i(S)=0\quad\text{if }i\notin S,
\qquad
\sum_{i\in N}y_i(S)=v(S),
\qquad
y_i(N)=x_i,
\end{equation}
and set $y_i(\varnothing)=0$. Define payments as changes in these
accounts:
\begin{equation}
f_i^y(S,T)=y_i(T)-y_i(S).
\label{eq:potentialflow}
\end{equation}
Along every path from $S$ to $N$, payments to player $i$ sum to
$x_i-y_i(S)$. Entry timing therefore does not affect the player's
total payment.

Let $\cI_{\rm loc}(v;x)$ denote the set of local, budget-balanced,
voluntary flows whose terminal allocation is $x$. Equip the flow space
with the Euclidean norm
\[
\|\mathbf f\|_2^2
:=
\sum_{i\in N}\sum_{(S,S\cup\{j\})\in\cE}
f_i(S,S\cup\{j\})^2.
\]

To state the invariance properties of the selected flow, consider a
permutation $\pi$ of the player set. The relabeled game, target
allocation, and flow are defined by
\[
(\pi v)(S):=v\bigl(\pi^{-1}(S)\bigr),
\qquad
(\pi x)_i:=x_{\pi^{-1}(i)},
\]
and
\[
(\pi\mathbf f)_i(S,S\cup\{j\})
:=
f_{\pi^{-1}(i)}
\left(
\pi^{-1}(S),
\pi^{-1}(S\cup\{j\})
\right).
\]
Thus, $\pi\mathbf f$ relabels coalitions, entrants, and payment
recipients in the same way.

\begin{proposition}[Target implementation and canonical redesign]
\label{prop:implementation}
Every efficient allocation $x$ admits a local, budget-balanced
voluntary flow. In particular, the flow in
\eqref{eq:potentialflow} implements $x$ and makes every consent
constraint bind.

Among all local voluntary flows that implement $x$, there is a unique
solution
\begin{equation}
\mathbf f^*(v,x)
=
\arg\min_{\mathbf f\in\cI_{\rm loc}(v;x)}
\frac12\|\mathbf f-\partial v\|_2^2.
\label{eq:projection}
\end{equation}
The selected flow is equivariant under player relabeling and positively
homogeneous. More precisely, for every permutation $\pi$ of $N$ and
every scalar $a>0$,
\begin{equation}
\mathbf f^*(\pi v,\pi x)
=
\pi\mathbf f^*(v,x),
\qquad
\mathbf f^*(av,ax)
=
a\mathbf f^*(v,x).
\label{eq:projection-invariance}
\end{equation}
If the Shapley flow is voluntary and $x=\phi(v)$, no redesign is
needed, and
$
\mathbf f^*(v,\phi(v))=\partial v.
$
\end{proposition}

The first identity in \eqref{eq:projection-invariance} means that the
selected payment rule does not depend on the names assigned to the
players. Relabeling the players, the game, and the target allocation
simply relabels the optimal payments. In particular, if a permutation
leaves both $v$ and $x$ unchanged, it also leaves
$\mathbf f^*(v,x)$ unchanged. The redesign therefore does not break
symmetries already present in the economic environment. The second identity means that the selected rule does not depend on
the unit in which coalition values and payments are measured.

The first part of Proposition~\ref{prop:implementation} is a broad
benchmark. With unrestricted signed transfers, voluntary entry places
no additional restriction on the set of efficient terminal
allocations. Cooperative criteria such as the core can select $x$,
while the flow determines how that allocation is delivered. The
second part addresses payment multiplicity. It treats the Shapley flow
as the status quo and changes it only as much as needed to satisfy
participation.

For a Shapley target, define
\[
D^{\rm Sh}(v)
:=
\|\mathbf f^*(v,\phi(v))-\partial v\|_2.
\]
This distance is a reduced-form measure of the payment adjustment
needed to preserve Shapley payoffs while supporting voluntary entry.
By Theorem~\ref{thm:main}, $D^{\rm Sh}(v)>0$ for every nonadditive
convex game. The distance is zero whenever the Shapley flow is already
voluntary, including every submodular game.

\subsection{How the redesign works in symmetric games}
\label{sec:cardinality}

The general minimum-redesign problem has a separate payment vector on
every edge of the coalition lattice. Even after imposing locality, its
dimension grows exponentially with the number of players. Symmetry
makes the economic adjustment much more transparent. 

Suppose that
coalition worth depends only on coalition size:
\begin{equation}
    v(S)=h(|S|),
    \qquad
    d_k:=h(k+1)-h(k),
    \qquad k=0,\ldots,n-1.
    \label{eq:cardinality-increments}
\end{equation}
Here $d_k$ is the value created by the $(k+1)$st entrant. The game is
submodular when $(d_k)$ is nonincreasing and convex when $(d_k)$ is
nondecreasing.

Because the game and the Shapley target are invariant under player
relabeling, Proposition~\ref{prop:implementation} implies that the
selected flow has the same symmetry. On every edge leaving a coalition
of size $k$, the selected flow therefore pays the entrant a common
amount $p_k$ and each of the $k$ incumbents a common amount $q_k$.
Edge-by-edge budget balance requires
\begin{equation}
    p_k+kq_k=d_k,
    \qquad\text{or equivalently}\qquad
    p_k=d_k-kq_k.
    \label{eq:cardinality-budget}
\end{equation}
The variable $q_k$ is the transfer to each incumbent when the coalition
expands from size $k$ to size $k+1$. A positive $q_k$ shifts some of the
current marginal surplus away from the entrant and toward players who
entered earlier. This reduces the advantage of waiting for a later
entry position. A negative $q_k$ moves payments in the opposite
direction. No incumbent exists at $k=0$, so we set $q_0=0$ and budget
balance fixes $p_0=d_0$.

Given $q=(q_0,q_1,\ldots,q_{n-1})$ with $q_0=0$, consider a player who
enters after exactly $k$ predecessors. At entry she receives
$p_k=d_k-kq_k$. At every subsequent expansion from size $\ell$ to
size $\ell+1$, she is an incumbent and receives $q_\ell$. Her total
payment along the remainder of the path is therefore
\begin{equation}
    r_k(q)
    :=d_k-kq_k+\sum_{\ell=k+1}^{n-1}q_\ell.
    \label{eq:entrypayoff}
\end{equation}
Thus $r_k(q)$ is not merely the payment made at the moment of entry. It
is the player's total path payment conditional on occupying entry
position $k+1$.

If the player defers after $k$ predecessors have entered, another outsider enters first. Under the full-consent continuation, each later entry position $k+2,\ldots,n$ is equally likely. Proposition~\ref{prop:dynamic-equivalence} therefore reduces full voluntariness to
\begin{equation}
    r_k(q)
    \geq
    \frac{1}{n-k-1}\sum_{\ell=k+1}^{n-1}r_\ell(q),
    \qquad k=0,\ldots,n-2.
    \label{eq:cardconstraints}
\end{equation}
The constraint at level $k$ says that immediate acceptance must be at least as attractive as deferral followed by the full-consent continuation.

\begin{proposition}[Cardinality games]
\label{prop:cardinality}
For a cardinality game $v(S)=h(|S|)$, the closest voluntary flow that
implements the Shapley allocation is determined by the unique solution
to
\begin{equation}
    \min_{q_1,\ldots,q_{n-1}}
    \frac12\sum_{k=1}^{n-1}
    \binom nk(n-k)k(k+1)q_k^2
    \label{eq:cardobjective}
\end{equation}
subject to \eqref{eq:cardconstraints}. The corresponding entrant
payment is $p_k^*=d_k-kq_k^*$.

If $h$ is concave, then $q_k^*=0$ for every $k$, so the selected flow
is the Shapley flow.
\end{proposition}

To understand the objective, note that there are
$\binom nk(n-k)$ edges leaving level $k$. Relative to the Shapley flow,
the entrant's payment changes by $-kq_k$, while each of the $k$
incumbents receives $q_k$. The squared adjustment on one edge is
therefore
\[
    (-kq_k)^2+kq_k^2=k(k+1)q_k^2.
\]
This yields \eqref{eq:cardobjective}. The reduction replaces an
exponential-dimensional flow problem with a quadratic program in
$n-1$ variables and $n-1$ participation constraints. Economically, a binding participation constraint identifies a stage at which the selected transfers make the player indifferent between entering immediately and deferring to a later position. A slack constraint indicates that immediate entry is strictly preferred at that stage.

\paragraph{Two players.}
\label{subsec:cardinality-two}

With two players, the position-contingent path payments are
\[
    r_0=d_0+q_1,
    \qquad
    r_1=d_1-q_1.
\]
There is one participation constraint, $r_0\geq r_1$, or
$q_1\geq\frac{d_1-d_0}{2}$. 
Since the objective is a positive multiple of $q_1^2$, the closest
feasible adjustment is
\[
    q_1^* = \max\left\{0,\frac{d_1-d_0}{2}\right\}.
\]
If the game is convex, $d_1\geq d_0$, and the constraint binds. Hence
\[
    r_0(q^*)=r_1(q^*)
    =\frac{d_0+d_1}{2}
    =\frac{h(2)}{2}.
\]
The adjustment exactly removes the gain from moving from the first to
the second entry position.

\paragraph{Three players.}
\label{subsec:cardinality-three}

With three players, the total path payments associated with the three
possible entry positions are
\begin{equation*}
 r_0(q)=d_0+q_1+q_2,
 \qquad
 r_1(q)=d_1-q_1+q_2,
 \qquad
 r_2(q)=d_2-2q_2.
 \label{eq:cardinality-three-payoffs}
\end{equation*}
The participation constraint at the empty coalition and the constraint
after one player has entered are, respectively,
\begin{equation}
 r_0(q)\geq\frac{r_1(q)+r_2(q)}{2},
 \qquad
 r_1(q)\geq r_2(q).
 \label{eq:cardinality-three-IC-payoffs}
\end{equation}
After dropping an irrelevant common factor, the objective is
\[
 6q_1^2+9q_2^2.
\]

Now suppose that the game is convex, and define the successive
increases in marginal contribution by
\begin{equation*}
 u:=d_1-d_0\geq0,
 \qquad
 w:=d_2-d_1\geq0.
 \label{eq:cardinality-three-increments}
\end{equation*}
In terms of $u$ and $w$, the participation constraints in
\eqref{eq:cardinality-three-IC-payoffs} become
\begin{equation}
 q_1+q_2\geq\frac{2u+w}{3},
 \qquad
 -q_1+3q_2\geq w.
 \label{eq:cardinality-three-IC-uw}
\end{equation}
The unique solution is
\begin{equation}
(q_1^*,q_2^*)
=
\begin{cases}
\displaystyle
\left(\frac{2u+w}{5},\frac{2(2u+w)}{15}\right),
&\displaystyle w\leq\frac{u}{2},\\[4mm]
\displaystyle
\left(\frac{u}{2},\frac{u+2w}{6}\right),
&\displaystyle w>\frac{u}{2}.
\end{cases}
\label{eq:cardinality-three-solution-uw}
\end{equation}
The threshold $w=u/2$ determines whether the second participation
constraint affects the closest feasible flow. To see this, first ignore the second constraint in
\eqref{eq:cardinality-three-IC-uw}. Minimizing
$6q_1^2+9q_2^2$ subject to
$q_1+q_2\geq(2u+w)/3$ gives
$(q_1,q_2) =
 \left(\frac{2u+w}{5},\frac{2(2u+w)}{15}\right)$. 
At this point,
$-q_1+3q_2=\frac{2u+w}{5}$, 
so the second constraint is satisfied if and only if
$\frac{2u+w}{5}\geq w$, that is $ w\leq\frac{u}{2}$.

If instead $w>u/2$, this projection violates the second constraint, and
the optimum occurs where both constraints bind. Solving
$q_1+q_2=\frac{2u+w}{3}$ and  $-q_1+3q_2=w$
gives the second line of
\eqref{eq:cardinality-three-solution-uw}.

The two regions have different economic interpretations. If
$w<u/2$, the final increase in marginal contribution is relatively
small. Only the participation constraint at the empty coalition binds and gives
$r_0(q^*)=\frac{r_1(q^*)+r_2(q^*)}{2}$, 
whereas the constraint after one player has entered is strictly slack:
$r_1(q^*)>r_2(q^*)$. 
Thus a player at the empty coalition is indifferent between entering
first and giving up the first position. Once one player has entered,
however, she strictly prefers to enter second rather than wait to enter
last. The minimum-distance rule corrects the initial holdout incentive
without unnecessarily equalizing the payoffs from the second and final
positions.

If $w>u/2$, the increase from the second to the final marginal
contribution is sufficiently large that both participation constraints
bind, giving
$r_1(q^*)=r_2(q^*)$ and 
 $r_0(q^*)=\frac{r_1(q^*)+r_2(q^*)}{2}$.
It follows that
$ r_0(q^*)=r_1(q^*)=r_2(q^*)
 =\frac{d_0+d_1+d_2}{3}
 =\frac{h(3)}{3}$, 
hence the player receives the same total path payment whether she
enters first, second, or last. In this region, the selected flow fully
insures the player against her position in the entry order. At the boundary $w=u/2$, both constraints bind and the two formulas in
\eqref{eq:cardinality-three-solution-uw} coincide. This distinction also
explains why the closest voluntary flow need not be an all-binding
account-based implementation. When the second participation constraint
is slack, forcing equality across all three entry positions would alter
the Shapley flow more than voluntary entry requires.

\section{When the payment institution selects the allocation}
\label{sec:equalresidual}

The previous sections fix a terminal allocation and ask how to implement
it. We now reverse the order of the problem. We first restrict the form
of transition payments and then let voluntary entry and the
minimum-adjustment criterion determine the terminal allocation.

On an edge $S\to S\cup\{j\}$, let $p_j(S)$ denote the payment to the
entrant. Under \emph{equal residual sharing}, the part of the transition
surplus not paid to entrant $j$ is divided equally among the incumbents.
Thus, for every nonempty $S$ and $j\notin S$,
\begin{equation}
 f_i(S,S\cup\{j\})
 =
 \begin{cases}
 p_j(S), & i=j,\\[1mm]
 \displaystyle
 q_j(S):=\frac{\Delta_jv(S)-p_j(S)}{|S|}, & i\in S,\\[3mm]
 0, & i\notin S\cup\{j\}.
 \end{cases}
 \label{eq:ER}
\end{equation}
On an edge leaving the empty coalition there are no incumbents, so
budget balance requires
\begin{equation}
 p_j(\varnothing)=v(\{j\}).
 \label{eq:ER-empty}
\end{equation}
The Shapley flow belongs to this class: it sets
$p_j(S)=\Delta_jv(S)$ and therefore $q_j(S)=0$. Equal residual sharing
does not require redistribution when marginal-contribution pricing is
already voluntary. It only specifies how a departure from that
benchmark is distributed among incumbents.

\paragraph{A feasible equal-division benchmark.}
The equal-residual restriction always admits a voluntary flow when
signed payments are allowed. For every coalition $S$, define player
$i$'s equal-division account by
\begin{equation}
 y_i^{\rm ED}(S)
 :=
 \begin{cases}
 \displaystyle\frac{v(S)}{|S|},
     & i\in S\text{ and }S\neq\varnothing,\\[2mm]
 0,  & i\notin S\text{ or }S=\varnothing.
 \end{cases}
 \label{eq:ER-ED-accounts}
\end{equation}
The associated account-change flow is
\begin{equation}
 f_i^{\rm ED}(S,T)
 :=y_i^{\rm ED}(T)-y_i^{\rm ED}(S)
 \qquad\text{for every forward edge }(S,T).
 \label{eq:ER-ED-flow}
\end{equation}
This is an exact flow because it is the discrete gradient of the
coalition accounts $y^{\rm ED}$. Along every path from $S$ to $N$, the
payments to player $i$ telescope to $y_i^{\rm ED}(N)-y_i^{\rm ED}(S)$. 
Hence the player's remaining total payment does not depend on the
subsequent entry order, and every consent constraint binds.

The same construction also satisfies equal residual sharing. Let
$T=S\cup\{j\}$ and $k=|S|\geq1$. The entrant receives
\begin{equation}
 p_j^{\rm ED}(S)=\frac{v(T)}{k+1},
 \label{eq:ER-ED-entrant}
\end{equation}
and each incumbent receives
\begin{equation}
 q_j^{\rm ED}(S)
 =
 \frac{v(T)}{k+1}-\frac{v(S)}{k}.
 \label{eq:ER-ED-incumbent}
\end{equation}
Indeed,
\[
 \Delta_jv(S)-p_j^{\rm ED}(S)
 =
 k\left[
 \frac{v(T)}{k+1}-\frac{v(S)}{k}
 \right]
 =kq_j^{\rm ED}(S).
\]
Thus the part of the transition surplus not paid to the entrant is
exactly divided among the $k$ incumbents. At the grand coalition, every
player holds the account $v(N)/n$, so the terminal allocation is the equal division
\begin{equation}
 \ED(v):=\frac{v(N)}{n}\mathbf 1.
 \label{eq:equal-division}
\end{equation}

Let $\cI^{\rm ER}(v)$ denote the voluntary flows satisfying
\eqref{eq:ER} and \eqref{eq:ER-empty}. Define the closest equal-residual
flow and its terminal allocation by
\begin{equation}
 \mathbf f^{\rm ER,*}(v)
 :=
 \arg\min_{\mathbf f\in\cI^{\rm ER}(v)}
 \frac12\|\mathbf f-\partial v\|_2^2,
 \qquad
 \Psi^{\rm ER}(v)
 :=
 \phi_{\mathbf f^{\rm ER,*}(v)}.
 \label{eq:ERvalue}
\end{equation}
Unlike the problem in Proposition~\ref{prop:implementation}, no target
allocation is imposed in \eqref{eq:ERvalue}. The payment convention,
the voluntary-entry constraints, and the minimum-adjustment criterion
jointly select the allocation.

\paragraph{Player types.}
Recall the relabeling operation from
Section~\ref{sec:implementation}. The automorphism group of the game is
\begin{equation*}
 \Aut(v):=\{\pi:\pi v=v\}. 
\end{equation*}
Players $i$ and $j$ have the same \emph{type} if some
$\pi\in\Aut(v)$ maps $i$ to $j$. Equivalently, player types are the
orbits of $\Aut(v)$ on $N$. Players of the same type are economically
interchangeable: relabeling them leaves every coalition value
unchanged.

\begin{proposition}[The equal-residual value]
\label{prop:ER}
The program in \eqref{eq:ERvalue} has a unique solution for every game.
As in Proposition~\ref{prop:implementation}, the selected flow is
equivariant under player relabeling and positively homogeneous:
\begin{equation}
 \mathbf f^{\rm ER,*}(\pi v)
 =\pi\mathbf f^{\rm ER,*}(v),
 \qquad
 \mathbf f^{\rm ER,*}(av)
 =a\mathbf f^{\rm ER,*}(v)
 \quad(a>0).
 \label{eq:ER-invariance}
\end{equation}
Consequently, $\Psi^{\rm ER}$ is efficient and satisfies
\begin{equation}
 \Psi^{\rm ER}(\pi v)=\pi\Psi^{\rm ER}(v),
 \qquad
 \Psi^{\rm ER}(av)=a\Psi^{\rm ER}(v).
 \label{eq:ER-value-invariance}
\end{equation}
In particular, players of the same type receive the same payoff. If the
Shapley flow is voluntary, then
\[
 \mathbf f^{\rm ER,*}(v)=\partial v
 \qquad\text{and}\qquad
 \Psi^{\rm ER}(v)=\phi(v).
\]

Suppose, in addition, that $\Aut(v)$ has exactly two orbits, denoted by
$A$ and $B$, and that the Shapley value distinguishes the two types:
\begin{equation}
\phi_i(v)\neq\phi_j(v) \quad \text{for } \ i\in A,\ j\in B. 
 \label{eq:ER-two-types-nondegenerate}
\end{equation}
Then there is a unique scalar $\alpha^{\rm ER}(v)\in\R$ such that
\begin{equation}
 \Psi^{\rm ER}(v)
 =
 \alpha^{\rm ER}(v)\phi(v)
 +\bigl[1-\alpha^{\rm ER}(v)\bigr]\ED(v).
 \label{eq:mixture}
\end{equation}
\end{proposition}

The invariance properties have the same meaning as in
Proposition~\ref{prop:implementation}. Relabeling the players merely
relabels the selected payments, while multiplying all coalition values
by a positive factor multiplies every selected payment by that factor.
If $\pi\in\Aut(v)$, then $\pi v=v$, and uniqueness implies that the
selected flow is invariant under $\pi$. Players in the same orbit
therefore receive equal terminal payoffs.

The two-type representation in \eqref{eq:mixture} is geometric. An
allocation that is constant on $A$ and on $B$ is described by two
numbers, while efficiency imposes one linear restriction. The set of
type-symmetric efficient allocations is therefore an affine line. The
Shapley and equal-division allocations are distinct points on that line
under \eqref{eq:ER-two-types-nondegenerate}, so they span it. What the
payment problem selects is the location $\alpha^{\rm ER}(v)$ on this
line. If $\alpha^{\rm ER}(v)\in[0,1]$, the selected allocation lies
between the Shapley and equal-division allocations. With three or more
player types, the type-symmetric efficient set generally has more than
one degree of freedom, so a single mixing weight need not summarize the
correction. 

\subsection{The glove game}
\label{subsec:ER-glove}

Player 1 owns a left glove and players 2 and 3 each own a right glove. A
coalition has value one if and only if it contains player 1 and at least
one right-glove player. The two player types are $A=\{1\}$, $B=\{2,3\}$. 
The Shapley allocation is
\[
 \phi(v)=\left(\frac23,\frac16,\frac16\right),
\]
but the Shapley flow is not voluntary. The solution of the equal-residual program, derived in Appendix~\ref{app:glove-calculations}, gives
\begin{equation}
 \Psi^{\rm ER}(v)
 =
 \left(\frac7{12},\frac5{24},\frac5{24}\right)
 =
 \frac34\phi(v)+\frac14\ED(v).
 \label{eq:glove}
\end{equation}
The scarce left glove retains a premium. Voluntary entry reduces that
premium because the right-glove players must be willing to occupy early
entry positions at which their marginal-contribution payment is low.

\paragraph{Nonnegative equal-residual payments.}
The preceding program permits signed individual payments. If negative
payments are infeasible, impose
\begin{equation}
 p_j(S)\geq0,
 \qquad
 q_j(S)\geq0
 \qquad(S\neq\varnothing,\ j\notin S).
 \label{eq:ER-nonnegative}
\end{equation}
Nonnegative feasibility is not automatic. A convenient sufficient
condition is
\begin{equation}
 v(S)\geq0
 \quad\text{for every }S\subseteq N, \ \text{ and}
 \quad
 \frac{v(S\cup\{j\})}{|S|+1}
 \geq
 \frac{v(S)}{|S|}
 \quad(S\neq\varnothing,\ j\notin S).
 \label{eq:ER-nonnegative-sufficient}
\end{equation}
The second condition says average coalition worth does not fall
when the coalition expands.

The equal-division account flow shows why these conditions are
sufficient. Its entrant payment is
$
 p_j^{\rm ED}(S)
 =\frac{v(S\cup\{j\})}{|S|+1}\geq0,
$
and each incumbent receives
$
 q_j^{\rm ED}(S)
 =
 \frac{v(S\cup\{j\})}{|S|+1}
 -\frac{v(S)}{|S|}
 \geq0.
$
On an edge leaving the empty coalition,
$p_j^{\rm ED}(\varnothing)=v(\{j\})\geq0$. Hence the same exact flow that
establishes nonemptiness of $\cI^{\rm ER}(v)$ also gives a nonnegative
voluntary implementation under
\eqref{eq:ER-nonnegative-sufficient}. The closest nonnegative
equal-residual flow therefore exists and is unique.

For the glove game, imposing the restrictions \eqref{eq:ER-nonnegative}  yields the allocation (see Appendix~\ref{app:glove-calculations})
\begin{equation}
 \Psi_+^{\rm ER}(v)
 =
 \left(\frac{12}{23},\frac{11}{46},\frac{11}{46}\right)
 =
 \frac{13}{23}\phi(v)+\frac{10}{23}\ED(v).
 \label{eq:gloveplus}
\end{equation}
The nonnegativity restriction requires a larger movement toward equal
division in this example. We remark that condition
\eqref{eq:ER-nonnegative-sufficient} is sufficient but not necessary.
For example, an additive monotone game may violate nondecreasing average
worth even though its Shapley flow, with zero residual payments, is
already a nonnegative voluntary flow.

\subsection{Participation and coalitional stability}
\label{subsec:ER-core}

Equal residual sharing is a procedural fairness rule. It does not by
itself ensure that the resulting allocation belongs to the core.
Consider the three-player unanimity game
\[
 v(S)=1\quad\text{if }\{1,2\}\subseteq S,
 \qquad
 v(S)=0\quad\text{otherwise}.
\]
The game is convex. Players 1 and 2 form one type, while player 3 is a
null player of a second type. The core requires $x_3=0$, but the
equal-residual program gives
\begin{equation}
 \Psi^{\rm ER}(v)
 =
 \left(\frac{11}{24},\frac{11}{24},\frac1{12}\right)
 =
 \frac34\phi(v)+\frac14\ED(v).
 \label{eq:corecounter}
\end{equation}
Hence players 1 and 2 jointly receive
$11/12<v(\{1,2\})=1$. The payment process compensates the null player
for her position in coalition formation, but the productive pair can
object to financing that compensation.

This example identifies a boundary of the equal-residual rule. Local
procedural compensation can conflict with a coalition's outside option.
If coalitional stability is required, core constraints may be imposed
directly on the terminal allocation. Since $\phi_{\mathbf f}$ is linear
in $\mathbf f$, the restrictions for core membership, $\phi_{\mathbf f}(S)\geq v(S)$ for $S\subseteq N$, 
are linear in the flow. Adding them preserves convexity and uniqueness
of the projection whenever the core-constrained feasible set is
nonempty.

For two-type games, the mixture representation gives a more explicit
comparison between participation and core stability. Let
\begin{equation}
 e_v(S):=\frac{|S|}{n}v(N)
 \label{eq:equal-division-coalition-payoff}
\end{equation}
denote the amount coalition $S$ receives under equal division. Summing
\eqref{eq:mixture} over $S$ gives
\begin{equation}
 \Psi^{\rm ER}(v)(S)
 =
 e_v(S)
 +\alpha^{\rm ER}(v)
 \bigl[\phi(v)(S)-e_v(S)\bigr].
 \label{eq:ER-coalition-mixture}
\end{equation}
Suppose equal division falls short of the worth of $S$:
$v(S)>e_v(S)$. If
$\phi(v)(S)>e_v(S)$, the core constraint of coalition $S$ is equivalent
to (see Appendix \ref{app:ER-core-threshold})
\begin{equation}
 \alpha^{\rm ER}(v)
 \geq
 \frac{v(S)-e_v(S)}
      {\phi(v)(S)-e_v(S)}.
 \label{eq:ER-core-threshold}
\end{equation}
Thus a coalition with a strong outside option places a lower bound on
the weight assigned to the Shapley allocation. For a convex game, we have $ \phi(v)(S)\geq v(S) > e_v(S)$, hence the
denominator is automatically positive whenever equal division falls
short. The restriction in \eqref{eq:ER-core-threshold} is therefore
well-defined in the convex case.

If the game is convex and $\alpha^{\rm ER}(v)\in[0,1]$, coalitions for
which $v(S)\leq e_v(S)$ impose no additional restriction: both the
Shapley allocation and equal division satisfy their core inequality,
and so does every convex combination of the two. Consequently,
\begin{equation}
 \Psi^{\rm ER}(v)\in\Core(v)
 \quad\Longleftrightarrow\quad
 \alpha^{\rm ER}(v)
 \geq
 \underline\alpha_{\rm core}(v),
 \label{eq:ER-core-characterization}
\end{equation}
where
\begin{equation}
 \underline\alpha_{\rm core}(v)
 :=
 \max_{\substack{S\subseteq N:\\ v(S)>e_v(S)}}
 \frac{v(S)-e_v(S)}
      {\phi(v)(S)-e_v(S)},
 \label{eq:ER-core-global-threshold}
\end{equation}
with the maximum over an empty collection interpreted as zero. 

In the unanimity game above, coalition $S=\{1,2\}$ has
\[
 v(S)=1,
 \qquad
 e_v(S)=\frac23,
 \qquad
 \phi(v)(S)=1.
\]
Hence $\underline\alpha_{\rm core}(v)=1$. Core stability permits no
strict movement from the Shapley allocation toward equal division,
whereas the equal-residual payment problem selects
$\alpha^{\rm ER}(v)=3/4$. The example makes the tension precise:
voluntary entry calls for procedural compensation, while coalitional
stability limits how much of that compensation the productive coalition
can finance.

\section{Conclusion}

The random-order interpretation of the Shapley value combines two objects: an allocation of grand-coalition surplus and a payment path
that produces it. The paper separates them. We formulate a finite dynamic entry problem and show that sequential optimality of immediate acceptance is equivalent to linear inequalities in full-consent continuation values. This distinction matters because a desirable terminal allocation need not be supported by the payments implicit in its usual procedural interpretation.

For the Shapley flow, the relevant incentive is straightforward. A selected outsider compares her current marginal contribution with the option value of deferral. Submodularity discourages deferral. Convexity, by increasing marginal contributions, can create
holdout. As a result, every nonadditive convex game exhibits a sharp
contrast: the Shapley allocation belongs to the core, but the Shapley
payment path is not voluntary.

This conflict does not make the allocation unattainable. With signed
local transfers, every efficient target can be generated by a
budget-balanced voluntary flow. The minimum-distance criterion selects
one such flow by changing marginal-contribution pricing only as much as
participation requires. In symmetric cardinality games, the selected
adjustments can be described by a small number of incumbent transfers,
which identify the stages at which waiting incentives bind.

Equal residual sharing illustrates how a payment institution can also
select an allocation. The rule treats incumbents equally on each edge
without imposing equal terminal payoffs. Its closest voluntary flow
agrees with Shapley when no adjustment is needed. With two player types,
the resulting allocation is an endogenous affine combination of the
Shapley and equal-division allocations. The glove game shows how
nonnegativity restrictions can change that combination.

The unanimity example also shows the limit of the construction.
Payments that support voluntary entry can assign compensation to a
player who is unproductive in the characteristic function, thereby
violating the outside option of a productive coalition. Procedural
participation and core stability are therefore distinct requirements.
For two-type convex games, the lower bound on the Shapley weight makes
this tradeoff explicit.

Several questions remain. One is to characterize nonnegative voluntary
implementation beyond the sufficient condition used here. Another is
to determine when equal residual sharing and core constraints are
jointly feasible. A third is to enrich the entry protocol with discounting, permanent withdrawal, or endogenous beliefs. A fourth is to study alternative adjustment costs and payment restrictions. These extensions would clarify which conclusions
come from voluntary entry itself and which depend on the institution
used to redistribute transition surplus.

\appendix
\section{Proofs and calculations}

\subsection{Proof of Proposition \ref{prop:dynamic-equivalence}}

We first establish the equivalence of statements 1--3 by backward
induction on the number of outsiders $
m(S)=n-|S|$. Recall that $V_i(N)=U_i(N)=0$, and for $|S| = n-1$,  $V_i(S)=U_i(S)= f_i (S,N)$, since deferral is no
longer available.

Suppose that the flow is fully voluntary, as in statement 1. Fix a state $S$ and suppose $
V_i(T)=U_i(T)$ for every strict successor $T\supsetneq S$. It follows that
$A_i(S;V)=A_i(S;U)$ and 
$B_i(S;V)=B_i(S;U)$. 
Because the flow is fully voluntary, immediate acceptance is optimal
at $S$, so
$A_i(S;V)\ge B_i(S;V)$. 
The Bellman recursion for $V_i$ therefore becomes
\begin{align*}
V_i(S)
&=
\frac{1}{m(S)}A_i(S;V)
+
\frac{m(S)-1}{m(S)}B_i(S;V)\\
&=
\frac{1}{m(S)}A_i(S;U)
+
\frac{m(S)-1}{m(S)}B_i(S;U)\\
&=U_i(S),
\end{align*}
where the last equality follows from
\eqref{eq:U-convex-combination}. Thus statement 1 implies
statement 2.

Conversely, suppose that statement 2 holds, so that
$V_i(T)=U_i(T)$ 
for every $i\in N$ and $T\subseteq N$. Fix a coalition
$S$ with $i\notin S$ and $m(S)\ge 2$. Equality at the successor
states implies
$A_i(S;V)=A_i(S;U)$ and 
$B_i(S;V)=B_i(S;U)$. 
Using the Bellman recursion for $V_i$, the full-consent recursion for
$U_i$, and the equality $V_i(S)=U_i(S)$, we obtain
\begin{align*}
&\frac{1}{m(S)}
\max\{A_i(S;U),B_i(S;U)\}
+
\frac{m(S)-1}{m(S)}B_i(S;U)
 =
\frac{1}{m(S)}A_i(S;U)
+
\frac{m(S)-1}{m(S)}B_i(S;U).
\end{align*}
After cancelling the common terms, this gives
$\max\{A_i(S;U),B_i(S;U)\}=A_i(S;U)$, and hence
$A_i(S;U)\ge B_i(S;U)$. 
Therefore statement 2 implies statement 3.

We next show that statement 3 implies statement 1. Suppose 
$A_i(S;U)\ge B_i(S;U)$
at every decision state. We again proceed by backward induction.
The equality $V_i(T)=U_i(T)$ holds at terminal states. Suppose it
holds at every strict successor $T\supsetneq S$. Then
$A_i(S;V)=A_i(S;U)$ and 
$B_i(S;V)=B_i(S;U)$. 
Statement 3 therefore implies
$A_i(S;V)\ge B_i(S;V)$, 
so immediate acceptance is optimal at $S$. Moreover, the Bellman
recursion reduces to
\begin{align*}
V_i(S)
&=
\frac{1}{m(S)}A_i(S;V)
+
\frac{m(S)-1}{m(S)}B_i(S;V)\\
&=
\frac{1}{m(S)}A_i(S;U)
+
\frac{m(S)-1}{m(S)}B_i(S;U)\\
&=U_i(S).
\end{align*}
Thus immediate acceptance is optimal at every decision state, and the
flow is fully voluntary. 

It remains to establish the equivalence of statements 3 and 4.
Equation \eqref{eq:U-convex-combination} gives
\[
U_i(S)
=
\frac{1}{m(S)}A_i(S;U)
+
\frac{m(S)-1}{m(S)}B_i(S;U).
\]
Because $m(S)\ge 2$, we have
\begin{align*}
A_i(S;U)\ge B_i(S;U)
&\quad\Longleftrightarrow\quad
A_i(S;U)\ge U_i(S).
\end{align*}
The last inequality is \eqref{eq:consent}. Hence statements 3 and 4 are equivalent. Since each $U_i(S;\mathbf f)$ is linear in $\mathbf f$, the constraints \eqref{eq:consent} are linear, and the set of fully voluntary flows is a closed convex polyhedron.

\subsection{Proof of Proposition \ref{prop:superharmonicity}}
Let $m=m(S) = n-|S|$ and $i\notin S$. Under locality, $f_i(S,S\cup\{j\})=0$ whenever $j\neq i$ is an outsider. Equation \eqref{eq:bellman} therefore implies
\[
mU_i(S)=f_i(S,S\cup\{i\})+U_i(S\cup\{i\})+
\sum_{j\in N\setminus(S\cup\{i\})}U_i(S\cup\{j\}).
\]
Substitution into \eqref{eq:gap} gives \eqref{eq:gap-identity}; dividing by $m-1$ yields \eqref{eq:superharmonicity}. Proposition~\ref{prop:dynamic-equivalence} then gives the claimed characterization of full voluntariness.

\subsection{Proof of Proposition \ref{prop:voluntary}}
Under the Shapley flow, player $i$ is paid only when she enters. Immediate acceptance at $S$ gives $\Delta_i v(S)$ and no subsequent payment. Under the full-consent continuation after deferral, player $i$ enters at the random predecessor coalition $R_i^S$ and receives $\Delta_i v(R_i^S)$. Proposition~\ref{prop:dynamic-equivalence} therefore yields \eqref{eq:exacttest}.

If $v$ is submodular, $R_i^S\supseteq S$ implies $\Delta_i v(S)\ge\Delta_i v(R_i^S)$ pathwise, so all linear consent inequalities hold and immediate acceptance is sequentially optimal at every state. If $v$ is supermodular, the reverse comparison holds. Suppose $\Delta_i v(S)<\Delta_i v(T)$ for some $S\subseteq T\subseteq N\setminus\{i\}$. Conditional on deferring at $S$, there is positive probability that all players in $T\setminus S$ enter before $i$. Monotonicity then implies $\mathbb E^S[\Delta_i v(R_i^S)]>\Delta_i v(S)$. The linear constraint at $S$ fails, and Proposition~\ref{prop:dynamic-equivalence} implies that immediate acceptance is not optimal in the dynamic entry problem.

\subsection{Proof of Theorem \ref{thm:main}}
For a convex game, every permutation marginal vector belongs to the core \citep{Shapley1971,Ichiishi1981}. Since the core is convex and the Shapley value is the average of the permutation marginal vectors, the Shapley allocation also belongs to the core. If a convex game is nonadditive, some player's marginal contribution must change along a nested pair of coalitions. Supermodularity makes such change strictly positive. Proposition \ref{prop:voluntary} then gives a violated entry constraint. If the game is additive, each marginal contribution is independent of the predecessor coalition, so every consent inequality binds.

\subsection{Proof of Proposition \ref{prop:implementation}}

We first prove existence of a voluntary implementation. Choose
coalition accounts $y_i(S)$ satisfying \eqref{def:accounts}, i.e., 
$y_i(S)=0$ for $i\notin S$, $\sum_{i\in N}y_i(S)=v(S)$, $y_i(N)=x_i$, 
and $y_i(\varnothing)=0$. The flow in
\eqref{eq:potentialflow} is then local. It is also budget-balanced because,
for every forward edge $(S,T)$,
\[
\sum_{i\in N}f_i^y(S,T)
=
\sum_{i\in N}y_i(T)-\sum_{i\in N}y_i(S)
=
v(T)-v(S).
\]
Along every forward path from $S$ to $N$, payments telescope:
\[
\sum_t f_i^y(X_{t-1},X_t)
=
y_i(N)-y_i(S)
=
x_i-y_i(S).
\]
Hence the full-consent continuation payoff is
$U_i(S;\mathbf f^y)=x_i-y_i(S)$. 
If $T=S\cup\{i\}$, then
\[
f_i^y(S,T)+U_i(T;\mathbf f^y)
=
y_i(T)-y_i(S)+x_i-y_i(T)
=
U_i(S;\mathbf f^y).
\]
Every consent constraint therefore binds, and
$U_i(\varnothing;\mathbf f^y)=x_i$. Thus
$\cI_{\rm loc}(v;x)$ is nonempty.

Continuation payoffs are linear in the flow by backward induction in
\eqref{eq:bellman}. Consequently, the consent constraints are linear
inequalities in $\mathbf f$. Budget balance, locality, and the target
condition $\phi_{\mathbf f}=x$ are affine equalities. Therefore
$\cI_{\rm loc}(v;x)$ is a nonempty closed convex polyhedron. The
objective in \eqref{eq:projection} is coercive and strictly convex, so
it has a unique minimizer.

We next make the relabeling argument precise. For a permutation
$\pi:N\to N$ and a coalition $S$, write
$\pi(S):=\{\pi(i):i\in S\}$. Recall that the action of $\pi$ on a flow is defined by
\begin{equation}
(\pi\mathbf f)_i(S,S\cup\{j\})
:=
f_{\pi^{-1}(i)}
\left(
\pi^{-1}(S),
\pi^{-1}(S)\cup\{\pi^{-1}(j)\}
\right). \nonumber
\end{equation}
A flow $\mathbf f$ belongs to $\cI_{\rm loc}(v;x)$ if and only if
$\pi\mathbf f$ belongs to
$\cI_{\rm loc}(\pi v;\pi x)$. Indeed, relabeling preserves locality,
budget balance, the terminal allocation, and every consent inequality.
Moreover,
$\partial(\pi v)=\pi(\partial v)$,
and the Euclidean norm is unchanged by a permutation of coordinates:
$\|\pi\mathbf f-\partial(\pi v)\|_2 = \|\mathbf f-\partial v\|_2$. It follows that $\pi\mathbf f^*(v,x)$ solves the optimization problem
associated with $(\pi v,\pi x)$. Uniqueness therefore gives
$\mathbf f^*(\pi v,\pi x)=\pi\mathbf f^*(v,x)$. 
Finally, fix $a>0$. A flow $\mathbf f$ belongs to
$\cI_{\rm loc}(v;x)$ if and only if $a\mathbf f$ belongs to
$\cI_{\rm loc}(av;ax)$. This follows because budget balance, the
terminal allocation, continuation payoffs, and the consent gaps all
scale by $a$. Since
$\partial(av)=a\partial v$ and
$\|a\mathbf f-a\partial v\|_2^2 = a^2\|\mathbf f-\partial v\|_2^2$, 
uniqueness again yields
$\mathbf f^*(av,ax)=a\mathbf f^*(v,x)$. 
Finally, if $x=\phi(v)$ and the Shapley flow is voluntary, then
$\partial v\in\cI_{\rm loc}(v;\phi(v))$ and attains objective value
zero. It is therefore the unique minimizer.

\subsection{Proof of Proposition \ref{prop:cardinality}}

Because both the cardinality game $v(S)=h(|S|)$ and its Shapley target
are invariant under every permutation of the player set, the relabeling
equivariance and uniqueness established in
Proposition~\ref{prop:implementation} imply that the selected flow is
symmetric. Consequently, on every edge leaving a coalition of size
$k$, the entrant receives a common payment $p_k$ and each of the $k$
incumbents receives a common payment $q_k$. Edge-by-edge budget balance
then gives
\[
 p_k+kq_k=d_k,
 \qquad\text{hence}\qquad
 p_k=d_k-kq_k.
\]
Since there is no incumbent when $k=0$, we set $q_0=0$ and obtain
$p_0=d_0$.

Consider a player who enters after exactly $k$ predecessors. She first
receives the entrant payment $p_k$. At every subsequent expansion from
coalition size $\ell$ to size $\ell+1$, where
$\ell=k+1,\ldots,n-1$, she is an incumbent and receives $q_\ell$.
Her total path payment conditional on entry at level $k$ is therefore
\[
 p_k+\sum_{\ell=k+1}^{n-1}q_\ell
 =d_k-kq_k+\sum_{\ell=k+1}^{n-1}q_\ell
 =r_k(q),
\]
which proves \eqref{eq:entrypayoff}. Suppose now that the player is offered entry after $k$ predecessors have entered and defers. Under the universal-acceptance continuation, conditional on the player still being outside, her eventual position is uniform over the remaining levels $k+1,\ldots,n-1$. Thus the full-consent value of deferral is
\[
 \frac{1}{n-k-1}
 \sum_{\ell=k+1}^{n-1}r_\ell(q).
\]
Immediate entry is voluntary if and only if $r_k(q)$ is at least this
quantity. Applying this comparison at every level
$k=0,\ldots,n-2$ gives precisely the constraints in
\eqref{eq:cardconstraints}. The constraint at $k=n-1$ is omitted
because the only remaining outsider is the last entrant, whose consent
constraint binds automatically.

It remains to reduce the distance objective. There are
$\binom{n}{k}$ coalitions of size $k$, and each has $n-k$ outgoing
edges. Hence there are $\binom{n}{k}(n-k)$ edges leaving level $k$. On each such edge, the Shapley flow pays the
entrant $d_k$ and pays every incumbent zero. Under the symmetric flow,
the entrant instead receives $p_k=d_k-kq_k$, a change of $-kq_k$, and
each of the $k$ incumbents receives $q_k$. The squared adjustment on one
edge is therefore
\[
 (p_k-d_k)^2+kq_k^2
 =(-kq_k)^2+kq_k^2
 =k(k+1)q_k^2.
\]
Summing over all levels and retaining the constant $1/2$ from
\eqref{eq:projection} yields the objective in
\eqref{eq:cardobjective}.

The constraints in \eqref{eq:cardconstraints} are linear in
$q=(q_1,\ldots,q_{n-1})$, while every coefficient in the quadratic
objective is strictly positive. The reduced program is therefore a
strictly convex quadratic program over a closed convex feasible set.
Feasibility follows from Proposition~\ref{prop:implementation}, or
directly from the equal-account construction applied to the Shapley
target. Hence the program has a unique solution $q^*$, and budget
balance uniquely determines the associated entrant payments as
$p_k^*=d_k-kq_k^*$.

Finally, suppose that $h$ is concave. Then the increments are
nonincreasing: $d_0\geq d_1\geq\cdots\geq d_{n-1}$. 
At zero incumbent sharing, $q_k=0$ for every $k$, we have $r_k(0)=d_k$. 
Therefore, for each $k=0,\ldots,n-2$,
\[
 r_k(0)=d_k
 \geq
 \frac{1}{n-k-1}\sum_{\ell=k+1}^{n-1}d_\ell
 =
 \frac{1}{n-k-1}\sum_{\ell=k+1}^{n-1}r_\ell(0).
\]
Thus $q=0$ satisfies every voluntary-entry constraint. It also gives
objective value zero, the smallest possible value. Strict convexity
then implies that $q^*=0$ is the unique minimizer, and the selected flow
coincides with the Shapley flow.

\subsection{Proof of Proposition \ref{prop:ER}}
\label{app:proof-ER}

The accounts in \eqref{eq:ER-ED-accounts} generate the exact flow
\eqref{eq:ER-ED-flow}. For $T=S\cup\{j\}$ and $|S|=k\geq1$, the entrant
and incumbent payments are given by \eqref{eq:ER-ED-entrant} and
\eqref{eq:ER-ED-incumbent}. Moreover,
\[
 \Delta_jv(S)-p_j^{\rm ED}(S)
 =
 v(T)-v(S)-\frac{v(T)}{k+1}
 =
 k\left[\frac{v(T)}{k+1}-\frac{v(S)}{k}\right].
\]
Hence the flow satisfies equal residual sharing. Since payments are
account changes, they telescope along every path. The continuation
payoff from $S$ is $U_i(S;\mathbf f^{\rm ED}) = y_i^{\rm ED}(N)-y_i^{\rm ED}(S)$, and the calculation in the proof of
Proposition~\ref{prop:implementation} shows that every consent
constraint binds. Therefore $\cI^{\rm ER}(v)$ is nonempty for every
game.

Equal residual sharing and budget balance are affine restrictions on
the flow, and the consent conditions are linear inequalities. Thus
$\cI^{\rm ER}(v)$ is a nonempty closed convex polyhedron. The objective
in \eqref{eq:ERvalue} is coercive and strictly convex, so it has a
unique minimizer.

For relabeling, the action on flows is the one defined in
Section~\ref{sec:implementation}. A flow $\mathbf f$ belongs to
$\cI^{\rm ER}(v)$ if and only if $\pi\mathbf f$ belongs to
$\cI^{\rm ER}(\pi v)$. Relabeling preserves equal residual sharing,
budget balance, locality, and every consent inequality. In addition,
$\partial(\pi v)=\pi(\partial v)$ and 
$ \|\pi\mathbf f-\partial(\pi v)\|_2 = \|\mathbf f-\partial v\|_2$. 
Uniqueness therefore gives $ \mathbf f^{\rm ER,*}(\pi v) = \pi\mathbf f^{\rm ER,*}(v)$. 
 
Similarly, for $a>0$, multiplication by $a$ maps
$\cI^{\rm ER}(v)$ onto $\cI^{\rm ER}(av)$, while
$\partial(av)=a\partial v$ and 
 $\|a\mathbf f-a\partial v\|_2^2 = a^2\|\mathbf f-\partial v\|_2^2$. 
Hence
$\mathbf f^{\rm ER,*}(av)=a\mathbf f^{\rm ER,*}(v)$. 

Efficiency of $\Psi^{\rm ER}(v)$ follows from edge-by-edge budget
balance. The flow equivariance just established implies
$\Psi^{\rm ER}(\pi v)=\pi\Psi^{\rm ER}(v)$, 
and flow homogeneity gives
$\Psi^{\rm ER}(av)=a\Psi^{\rm ER}(v)$. 
 
If $\pi\in\Aut(v)$, then $\pi v=v$, and uniqueness implies
$\pi\mathbf f^{\rm ER,*}(v)=\mathbf f^{\rm ER,*}(v)$. Thus players in
the same orbit of $\Aut(v)$ receive the same terminal payoff. If the
Shapley flow is voluntary, it belongs to $\cI^{\rm ER}(v)$ and attains
objective value zero, proving Shapley consistency.

Finally, suppose $\Aut(v)$ has exactly two orbits $A$ and $B$. Every
$\Aut(v)$-invariant allocation has the form
\[
 x_i=
 \begin{cases}
 x_A,&i\in A,\\
 x_B,&i\in B.
 \end{cases}
\]
Efficiency imposes
$|A|x_A+|B|x_B=v(N)$, 
so invariant efficient allocations form an affine line. Both
$\phi(v)$ and $\ED(v)$ lie on this line. Condition
\eqref{eq:ER-two-types-nondegenerate} makes them distinct, and they
therefore span the line. Since $\Psi^{\rm ER}(v)$ is invariant and
efficient, the representation \eqref{eq:mixture} follows, with a unique
coefficient $\alpha^{\rm ER}(v)$.

\subsection{Glove-game calculations}
\label{app:glove-calculations}

We provide the reduced programs underlying
\eqref{eq:glove} and \eqref{eq:gloveplus}. Player 1 owns the left
glove, and players 2 and 3 own the right gloves. By symmetry between
players 2 and 3, the entrant payments can be represented by
\begin{align*}
 a&:=p_2(\{1\})=p_3(\{1\}),\\
 b&:=p_1(\{2\})=p_1(\{3\}),\\
 c&:=p_3(\{2\})=p_2(\{3\}),\\
 d&:=p_3(\{1,2\})=p_2(\{1,3\}),\\
 e&:=p_1(\{2,3\}).
\end{align*}
Entrant payments on edges leaving the empty coalition are zero. Equal
residual sharing determines all incumbent payments from
$(a,b,c,d,e)$.

Up to the conventional factor $1/2$, the squared distance from the
Shapley flow is
\begin{equation}
 Q^{\rm G}(a,b,c,d,e)
 =
 4(a-1)^2+4(b-1)^2+4c^2+3d^2
 +\frac32(e-1)^2.
 \label{eq:glove-objective-app}
\end{equation}
The variables $a$, $b$, and $e$ are compared with one because the
corresponding entrant completes a left--right pair. On the other hand, the
variables $c$ and $d$ are compared with zero. Substitution into the Bellman recursion gives five distinct
nontrivial consent constraints, after accounting for the symmetry
between players 2 and 3. Writing $z=(a,b,c,d,e)$, 
\begin{align*}
 &H_1(z)
  :=
 \frac23-\frac{2a}{3}-\frac b3-\frac d6-\frac e3
 \geq0,
 \quad H_2(z)
  :=
 \frac5{12}-\frac a6-\frac b3-\frac c2
 -\frac{5d}{12}-\frac e{12}
 \geq0,
 \\
 &H_3(z)
 :=
 \frac a2-\frac{3d}{4}
 \geq0,
  \quad
 H_4(z)
 :=
 \frac b2-\frac d4-\frac e2
 \geq0,
   \quad
 H_5(z)
 :=
 \frac14+\frac c2-\frac d2-\frac e4
 \geq0.
\end{align*}
The remaining consent constraints are symmetric copies or last-entry
constraints that bind. 

The terminal allocation generated by a symmetric equal-residual flow
with variables $z$ is
\begin{equation}
 \begin{aligned}
 \phi_{f_1}
 =
 \frac13-\frac a3+\frac b3-\frac d3+\frac e3, \qquad
 \phi_{f_2} = \phi_{f_3}
 =
 \frac13+\frac a6-\frac b6+\frac d6-\frac e6.
 \end{aligned}
 \label{eq:glove-allocation-app}
\end{equation}

\paragraph{Signed payments.}
With signed payments allowed, the reduced problem minimizes $Q^{\rm G}(z)/2$ subject to $H_m(z)\geq0$, $m=1,\ldots,5$. Its unique solution is 
$z^{\rm s} =
 \left(
 \frac12,\frac34,0,-\frac16,\frac13
 \right)$. 
 Substitution into \eqref{eq:glove-allocation-app} yields
$ \Psi^{\rm ER}(v) =  \left(\frac7{12},\frac5{24},\frac5{24}\right) = \frac34\phi(v)+\frac14\ED(v)$, 
which is \eqref{eq:glove}. 

For an optimality check, at $z^{\rm s}$, only $H_1$ binds; the remaining slacks are $H_2(z^{\rm s})=\frac18$, $H_3(z^{\rm s})=\frac38$ and $H_4(z^{\rm s})=H_5(z^{\rm s})=\frac14$. Let $F^{\rm G}=Q^{\rm G}/2$. Then
$
 \nabla F^{\rm G}(z^{\rm s})
 =3\nabla H_1(z^{\rm s})$.
Thus the Kuhn--Tucker conditions hold with multiplier $3$ on $H_1$
and zero multipliers on the other consent constraints. Strict
convexity of the objective implies that $z^{\rm s}$ is the unique
minimizer.

\paragraph{Nonnegative payments.}
We require every individual edge payment to be nonnegative. In the
reduced variables, nonnegativity implies
 $0\leq a\leq1$,
 $0\leq b\leq1$,
 $c=0$,
 $d=0$,
 $0\leq e\leq1$.
For example, the edge $\{2\}\to\{2,3\}$ creates no value and gives
entrant 3 the amount $c$ and incumbent 2 the amount $-c$; requiring
both payments to be nonnegative forces $c=0$. The same reasoning on
$\{1,2\}\to N$ and $\{1,3\}\to N$ forces $d=0$. After imposing $c=d=0$, the  solution is
 $z^+ = \left( \frac{11}{23},\frac{17}{23},0,0,\frac7{23} \right)$. 
At $z^+$ the consent constraint $H_1$ binds, while
$ H_2(z^+)=\frac3{46}$,
 $H_3(z^+)=\frac{11}{46}$,
 $H_4(z^+)=\frac5{23}$,
 $H_5(z^+)=\frac4{23}$.
The variables $a$, $b$, and $e$ are interior to their bounds. On the
reduced space $(a,b,e)$, stationarity is verified by
$ \nabla_{(a,b,e)}F^{\rm G}(z^+) = \frac{72}{23} \nabla_{(a,b,e)}H_1(z^+)$. 
With the binding restrictions $c=d=0$, this verifies the
Kuhn--Tucker conditions. 
Finally, \eqref{eq:glove-allocation-app} gives
$\Psi_+^{\rm ER}(v) = \left(\frac{12}{23},\frac{11}{46},\frac{11}{46}\right)
 = \frac{13}{23}\phi(v)+\frac{10}{23}\ED(v)$.

\subsection{Derivation of the core lower bound
\eqref{eq:ER-core-threshold}}
\label{app:ER-core-threshold}

Suppose that $v$ has two player types and that the equal-residual value
has the representation
\begin{equation}
 \Psi^{\rm ER}(v)
 =
 \alpha^{\rm ER}(v)\phi(v)
 +\bigl[1-\alpha^{\rm ER}(v)\bigr]\ED(v).
 \label{eq:ER-mixture-appendix}
\end{equation}
For $S\subseteq N$, let
$
 e_v(S):=\ED(v)(S)=\frac{|S|}{n}v(N)$. 
Summing \eqref{eq:ER-mixture-appendix} over the members of $S$ yields
\begin{align}
 \Psi^{\rm ER}(v)(S)
 &=
 \alpha^{\rm ER}(v)\phi(v)(S)
 +\bigl[1-\alpha^{\rm ER}(v)\bigr]e_v(S)\nonumber\\
 &=
 e_v(S)
 +\alpha^{\rm ER}(v)
 \bigl[\phi(v)(S)-e_v(S)\bigr].
 \label{eq:ER-mixture-coalition}
\end{align}
The core constraint associated with $S$ is $\Psi^{\rm ER}(v)(S)\geq v(S)$. 
By \eqref{eq:ER-mixture-coalition}, this is equivalent to
\begin{equation}
 \alpha^{\rm ER}(v)
 \bigl[\phi(v)(S)-e_v(S)\bigr]
 \geq
 v(S)-e_v(S).
 \label{eq:ER-core-rearranged}
\end{equation}

Suppose first that equal division violates the coalition's core
constraint, i.e. $v(S)>e_v(S)$, and suppose
$\phi(v)(S)>e_v(S)$. Dividing
\eqref{eq:ER-core-rearranged} by the positive denominator then gives  \eqref{eq:ER-core-threshold}. Thus the displayed lower bound
is not merely necessary for coalition $S$; under the stated positivity
condition, it is equivalent to that coalition's core constraint.

If $v$ is convex, $\phi(v)(S)\geq v(S)$ since the Shapley allocation belongs to the core. Hence the positivity condition $\phi(v)(S) - e_v(S) > 0$ holds whenever
$v(S)>e_v(S)$. Moreover, the coalition-specific threshold lies in $(0,1]$, since
\[
 0<v(S)-e_v(S)
 \leq
 \phi(v)(S)-e_v(S).
\]
Taking the maximum over all coalitions for which equal division falls
short gives \eqref{eq:ER-core-global-threshold}.

To obtain the equivalence in
\eqref{eq:ER-core-characterization}, additionally suppose
$\alpha^{\rm ER}(v)\in[0,1]$. If $v(S)\leq e_v(S)$, then equal division
satisfies coalition $S$'s core inequality. Convexity also implies that
the Shapley allocation satisfies it. Since
\eqref{eq:ER-mixture-appendix} is then a convex combination of these
two allocations, coalition $S$'s core inequality holds automatically.
The only potentially restrictive coalitions are therefore those with
$v(S)>e_v(S)$, and their joint requirement is exactly
\[
 \alpha^{\rm ER}(v)
 \geq
 \underline\alpha_{\rm core}(v).
\]

For the unanimity game $v=u_{\{1,2\}}$, take $S=\{1,2\}$. Then
$v(S)=1$, $e_v(S)=\frac23$ and $\phi(v)(S)=1$, so 
 $\underline\alpha_{\rm core}(v) = \frac{1-2/3}{1-2/3} =1$.
The equal-residual value has $\alpha^{\rm ER}(v)=3/4$, and therefore violates the core constraint of coalition $\{1,2\}$.

\subsection{Convex unanimity counterexample}
\label{app:unanimity-counterexample}

Consider the three-player unanimity game associated with
$T=\{1,2\}$:
\[
 u_{\{1,2\}}(S)
 :=
 \begin{cases}
 1, & \{1,2\}\subseteq S,\\
 0, & \text{otherwise}.
 \end{cases}
\]
The game is convex. Players 1 and 2 are symmetric, whereas player 3 is
a null player. The Shapley allocation is $\phi\bigl(u_{\{1,2\}}\bigr) = \left(\frac12,\frac12,0\right)$, and the core is
\[
 \Core\bigl(u_{\{1,2\}}\bigr)
 =
 \left\{
 (x_1,x_2,0):
 x_1+x_2=1,\ x_1\geq0,\ x_2\geq0
 \right\}.
\]
Indeed, coalition $\{1,2\}$ can generate the entire grand-coalition
worth, so its core constraint requires $x_1+x_2\geq1$. Efficiency and
the singleton constraint $x_3\geq0$ then imply $x_3=0$.

We now solve the equal-residual projection. Symmetry reduces the
entrant payments to five variables:
\begin{align*}
 &a:=p_2(\{1\})=p_1(\{2\}), \quad
 b:=p_3(\{1\})=p_3(\{2\}), \quad
 c:=p_1(\{3\})=p_2(\{3\}),\\
 &d:=p_3(\{1,2\}), \quad
 e:=p_2(\{1,3\})=p_1(\{2,3\}).
\end{align*}
All entrant payments on edges leaving the empty coalition are zero.
Equal residual sharing determines every incumbent payment from
$(a,b,c,d,e)$.

Up to the conventional factor $1/2$, the squared distance from the
Shapley flow is
\begin{equation}
 Q(a,b,c,d,e)
 =
 4(a-1)^2+4b^2+4c^2
 +\frac32d^2+3(e-1)^2.
 \label{eq:unanimity-objective-app}
\end{equation}
The variables $a$ and $e$ are compared with one because the associated
entries complete the productive pair. The variables $b$, $c$, and $d$
are compared with zero because the corresponding entrant has zero
marginal contribution.

Let $z:=(a,b,c,d,e)$. After accounting for the symmetry between players
1 and 2, the five distinct nontrivial consent constraints can be
written as
\begin{align} \label{eq:unanimity-app}
 &G_1(z) := \frac5{12}-\frac a2-\frac b3-\frac c6 -\frac d{12}-\frac{5e}{12}\geq0,
\quad
 G_2(z) := \frac16-\frac b3-\frac{2c}{3} -\frac d3-\frac e6\geq0,
\\
 &G_3(z) := \frac14+\frac c2-\frac{3e}{4}\geq0,
\quad
 G_4(z) := \frac a2-\frac d4-\frac e2\geq0,
\quad
 G_5(z) := \frac14+\frac b2-\frac d2-\frac e4\geq0. \nn
\end{align}
The states and entrants represented by these inequalities are derived
below. Constraints on edges entering the grand coalition bind
automatically. The equal-residual problem is therefore the strictly
convex quadratic program that minimizes $Q(z)/2$ subject to $G_m(z)\geq0$, $m=1,\ldots,5$. The unique solution is
\begin{equation}
 z^*=(a^*,b^*,c^*,d^*,e^*)
 =
 \left(
 \frac{13}{19},
 -\frac4{19},
 \frac3{76},
 -\frac8{57},
 \frac{41}{114}
 \right).
 \label{eq:unanimity-solution-app}
\end{equation}
At this solution, $G_1$ and $G_3$ bind. The Kuhn--Tucker multiplier
vector, for the constraints ordered as above and the objective $Q/2$,
is $(\lambda_1,\lambda_2,\lambda_3,\lambda_4,\lambda_5) =
 \left(\frac{48}{19},0,\frac{22}{19},0,0\right)$, and the remaining constraint slacks are $G_2(z^*)=G_4(z^*)=\frac{15}{76}$ and $G_5(z^*)=\frac18$. Verification for optimality is provided below.

The terminal allocation generated by a symmetric equal-residual flow
with variables $(a,b,c,d,e)$ is 
$\phi_{f_1}=\phi_{f_2} = \frac13-\frac b6+\frac c6-\frac d6+\frac e6$,
$ \phi_{f_3} = \frac13+\frac b3-\frac c3+\frac d3-\frac e3$. 
Substituting \eqref{eq:unanimity-solution-app} gives
\[
 \Psi^{\rm ER}\bigl(u_{\{1,2\}}\bigr)
  =
 \left(
 \frac{11}{24},
 \frac{11}{24},
 \frac1{12}
 \right)
 =
 \frac34\phi\bigl(u_{\{1,2\}}\bigr) +\frac14\ED\bigl(u_{\{1,2\}}\bigr).
\]
The allocation is efficient, but it does not belong to the core,
because coalition $\{1,2\}$ receives
\[
 \Psi^{\rm ER}_1+\Psi^{\rm ER}_2
 =
 \frac{11}{12} < 1 = u_{\{1,2\}}(\{1,2\}).
\]
The core violation is exactly the amount $1/12$ assigned to the null
player. Thus equal residual sharing may use procedural compensation to
support voluntary entry even when the productive coalition can object
to financing that compensation. In the notation of
Appendix~\ref{app:ER-core-threshold}, coalition $\{1,2\}$ requires
$\alpha^{\rm ER}\geq1$, whereas the equal-residual projection selects
$\alpha^{\rm ER}=3/4$.

\paragraph{Derivation of the consent inequalities.}
We now derive \eqref{eq:unanimity-app}. The notation
$G_m(z)$ refers to the reduced constraint functions in this example,
whereas $g_i(S;\mathbf f)$ denotes the general consent gap of player
$i$ at state $S$. The correspondence is
\begin{equation}
\begin{aligned}
 G_1(z)&=g_1(\varnothing;\mathbf f),
 &\qquad
 G_2(z)&=g_3(\varnothing;\mathbf f),\\
 G_3(z)&=g_1(\{3\};\mathbf f),
 &G_4(z)&=g_2(\{1\};\mathbf f),\\
 G_5(z)&=g_3(\{1\};\mathbf f).
\end{aligned}
\label{eq:unanimity-gap-correspondence}
\end{equation}
The constraints omitted from this list are either symmetric copies or
last-entry constraints.

Equal residual sharing first determines all incumbent payments from
the entrant payments $(a,b,c,d,e)$. Up to the symmetry between players
1 and 2, the nontrivial edge payments are
\begin{align*}
 \{1\}\to\{1,2\}:&
 &f_2&=a,            &f_1&=1-a,\\
 \{1\}\to\{1,3\}:&
 &f_3&=b,            &f_1&=-b,\\
 \{3\}\to\{1,3\}:&
 &f_1&=c,            &f_3&=-c,\\
 \{1,2\}\to N:&
 &f_3&=d,            &f_1=f_2&=-\frac d2,\\
 \{1,3\}\to N:&
 &f_2&=e,            &f_1=f_3&=\frac{1-e}{2}.
\end{align*}
For example, $\{1\}\to\{1,2\}$ creates one unit of value. Paying
entrant 2 the amount $a$ therefore leaves $1-a$ for incumbent 1. By
contrast, $\{1\}\to\{1,3\}$ and $\{3\}\to\{1,3\}$ create no value, so
a positive entrant payment must be offset by an equal negative payment
to the incumbent. The payments on symmetric edges are obtained by
interchanging players 1 and 2.

Applying the Bellman recursion \eqref{eq:bellman} backward from the
grand coalition gives, at the two-player coalitions,
\begin{align*}
 U_1(\{1,2\})=U_2(\{1,2\})&=-\frac d2,
 &U_3(\{1,2\})&=d,\\
 U_1(\{1,3\})=U_3(\{1,3\})&=\frac{1-e}{2},
 &U_2(\{1,3\})&=e,\\
 U_2(\{2,3\})=U_3(\{2,3\})&=\frac{1-e}{2},
 &U_1(\{2,3\})&=e.
\end{align*}
At singleton coalitions, the continuation values needed below are
\begin{align}
 U_1(\{2\})=U_2(\{1\})
 &=\frac{2a-d+2e}{4},
 \label{eq:unanimity-U-productive-partner}\\
 U_1(\{3\})=U_2(\{3\})
 &=\frac{1+2c+e}{4},
 \label{eq:unanimity-U-productive-null}\\
 U_3(\{1\})=U_3(\{2\})
 &=\frac{1+2b+2d-e}{4}.
 \label{eq:unanimity-U-null-productive}
\end{align}
For instance, $U_2(\{1\})
 =\frac12\left[
 a+U_2(\{1,2\})+U_2(\{1,3\})
 \right]
 =\frac12\left(a-\frac d2+e\right)$, 
which gives \eqref{eq:unanimity-U-productive-partner}. The other two
expressions follow from the same recursion. At the empty coalition,
\begin{align*}
 U_1(\varnothing)=U_2(\varnothing)
 =\frac{2-b+c-d+e}{6}, \qquad
 U_3(\varnothing)
 =\frac{1+b-c+d-e}{3}.
\end{align*}
For example, player 1 receives zero on the first transition from the
empty coalition. Conditional on the identity of the first entrant, her
continuation value is $U_1(\{1\})$, $U_1(\{2\})$, or
$U_1(\{3\})$, each with probability $1/3$. Substitution of the
singleton continuation values gives the displayed expression for
$U_1(\varnothing)$.

We can now apply the superharmonicity identity
\eqref{eq:gap-identity}. At the empty coalition, the consent gap of
productive player 1 is
\begin{align*}
 g_1(\varnothing;\mathbf f)
 =2U_1(\varnothing)-U_1(\{2\})-U_1(\{3\}) =\frac5{12}-\frac a2-\frac b3-\frac c6
   -\frac d{12}-\frac{5e}{12}
 =G_1(z).
\end{align*}
Player 2 has the same
constraint by symmetry. For null player 3,
\begin{align*}
 g_3(\varnothing;\mathbf f)
 =2U_3(\varnothing)-U_3(\{1\})-U_3(\{2\})
 =\frac16-\frac b3-\frac{2c}{3}-\frac d3-\frac e6
 =G_2(z).
\end{align*}
At a singleton coalition, only two outsiders remain. The consent gap is
therefore the difference between the entrant's continuation value at
the current state and her continuation value after the other outsider
enters. For productive player 1 after null player 3 has entered,
\begin{align*}
 g_1(\{3\};\mathbf f)
 =U_1(\{3\})-U_1(\{2,3\}) =\frac14+\frac c2-\frac{3e}{4}
 =G_3(z).
\end{align*}
The corresponding constraint
for player 2 is symmetric. For productive player 2 after productive
player 1 has entered,
\begin{align*}
 g_2(\{1\};\mathbf f)
 =U_2(\{1\})-U_2(\{1,3\})
 =\frac a2-\frac d4-\frac e2
 =G_4(z).
\end{align*}
Finally, for null player 3
after productive player 1 has entered,
\begin{align*}
 g_3(\{1\};\mathbf f)
 =U_3(\{1\})-U_3(\{1,2\})
 =\frac14+\frac b2-\frac d2-\frac e4
 =G_5(z),
\end{align*}
which completes \eqref{eq:unanimity-app}. The corresponding constraints
at $\{2\}$ follow by interchanging players 1 and 2. If two players are
already present, only one outsider remains, and the last entrant's
consent constraint binds automatically. This exhausts all distinct
consent constraints.

\paragraph{Verification of the minimizer.}
For completeness, we verify the Kuhn--Tucker conditions. Let $F(z):=\frac12Q(z)$
and write the Lagrangian as
$
\mathcal L(z,\lambda)
:=
F(z)-\sum_{m=1}^5\lambda_mG_m(z)$, $\lambda_m\geq0$. 
The sign convention reflects that the constraints are written as
$G_m(z)\geq0$.

At the candidate $z^*$ in
\eqref{eq:unanimity-solution-app}, direct substitution gives $G_1(z^*)=G_3(z^*)=0$, 
while
$G_2(z^*)=G_4(z^*)=\frac{15}{76}$ and $G_5(z^*)=\frac18$. 
Thus $z^*$ is feasible, and complementary slackness requires zero multipliers for the three slack constraints.

The gradient of the objective at $z^*$ is
$\nabla F(z^*) =
\left(
-\frac{24}{19},
-\frac{16}{19},
\frac3{19},
-\frac4{19},
-\frac{73}{38}
\right)$, 
while 
$\nabla G_1 =
\left(
-\frac12,
-\frac13,
-\frac16,
-\frac1{12},
-\frac5{12}
\right)
$, 
$\nabla G_3 =
\left(
0,0,\frac12,0,-\frac34
\right)$.  
Using the multipliers $(\lambda_1,\lambda_2,\lambda_3,\lambda_4,\lambda_5) = \left(\frac{48}{19},0,\frac{22}{19},0,0\right)$, we obtain
$\nabla F(z^*) =
\frac{48}{19}\nabla G_1
+
\frac{22}{19}\nabla G_3$, 
equivalently
$\nabla_z\mathcal L(z^*,\lambda^*)=0$. 
Primal feasibility, dual feasibility, complementary slackness, and
stationarity therefore hold. Since $F$ is strictly convex and the
feasible set is convex, these conditions are sufficient, and $z^*$ is
the unique minimizer.

%%%%%%%%%%%%%%%%%%%%%

\bibliographystyle{plainnat} 
\bibliography{fShapley}

%%%%%%%%%%%%%%%%%%%%%

\end{document}